\documentclass[usenatbib,usegraphicx,useAMS]{mn2e}

\newcommand{\microns}{$\mu$m}

\def\brg{Br\,$\gamma$}
\def\hei{He\,{\sc i}}
\def\nai{Na\,{\sc i}}

\def\mgii{Mg\,{\sc ii}}

\def\choh{CH$_{3}$OH}
\def\ga{\mathrel{\hbox{\rlap{\hbox{\lower4pt\hbox{$\sim$}}}\hbox{$>$}}}}
\def\la{\mathrel{\hbox{\rlap{\hbox{\lower4pt\hbox{$\sim$}}}\hbox{$<$}}}}
\def\msunyr{$M$ \mbox{$_{\normalsize\odot}$}\rm{yr}$^{-1}$}
\def\msun{$M$\mbox{$_{\normalsize\odot}$}}
\def\mstar{$M$\mbox{$_{\star}$}}

\def\lsun{$L$\mbox{$_{\normalsize\odot}$}}
\def\kms{\,km~s$^{-1}$}

\def\arcsec{$^{\prime \prime}$}
\def\arcmin{$^{\prime}$}

\def\hii{H{\sc ii}}

\newcommand{\fig}[1]{Fig.\ \ref{#1}}
\newcommand{\Fig}[1]{Figure \ref{#1}}

\title[The disk, envelope \& outflow of W33A]{The circumstellar disk, envelope, and
  bi-polar outflow of the Massive Young Stellar Object W33A}
\author[B. Davies et al.]{Ben Davies$^{1,2}$, Stuart
  L.\ Lumsden$^{1}$, Melvin G.\ Hoare$^{1}$, Ren\'{e}
  D.\ Oudmaijer$^{1}$, \newauthor Willem-Jan de Wit$^{1}$
  \\ $^{1}$School of Physics \& Astronomy, University of Leeds,
  Woodhouse Lane, Leeds LS2 9JT, UK\\ $^{2}$Chester F.\ Carlson Center
  for Imaging Science, Rochester Institute of Technology, 54 Lomb
  Memorial Drive, Rochester,\\ NY 14623, USA}
\begin{document}

\date{Accepted ... Received ...}

\pagerange{\pageref{firstpage}--\pageref{lastpage}} \pubyear{2009}

\maketitle

\label{firstpage}

\begin{abstract}
The Young Stellar Object (YSO) W33A is one of the best known examples
of a massive star still in the process of forming. Here we present
Gemini North ALTAIR/NIFS laser-guide star adaptive-optics assisted
$K$-band integral-field spectroscopy of W33A and its inner reflection
nebula. In our data we make the first detections of a
rotationally-flattened outer envelope and fast bi-polar jet of a
massive YSO at near-infrared wavelengths. The predominant spectral
features observed are \brg, H$_2$, and a combination of emission and
absorption from CO gas. We perform a 3-D spectro-astrometric analysis
of the line emission, the first study of its kind. We find that the
object's \brg\ emission reveals evidence for a fast bi-polar jet on
sub-milliarcsecond scales, which is aligned with the larger-scale
outflow. The hybrid CO features can be explained as a combination of
hot CO emission arising in a disk close to the central star, while
cold CO absorption originates in the cooler outer envelope. Kinematic
analysis of these features reveals that both structures are rotating,
and consistent with being aligned perpendicularly to both the ionised
jet and the large-scale outflow. Assuming Keplerian rotation, we find
that the circumstellar disk orbits a central mass of $\ga$10\msun,
while the outer envelope encloses a mass of $\sim$15\msun. Our results
suggest a scenario of a central star accreting material from a
circumstellar disk at the centre of a cool extended rotating torus,
while driving a fast bi-polar wind. These results therefore provide
strong supporting evidence for the hypothesis that the formation
mechanism for high-mass stars is qualitatively similar to that of
low-mass stars.


\end{abstract}


\section{Introduction}
It is still unclear how massive stars (\mstar $\ga$ 8\msun) are
formed. This is due in part to the rapid formation timescales of such
stars ($\la 10^{5}$yrs), which mean that the star can reach the
main-sequence (MS) while it is still heavily embedded in its natal
molecular cloud. Also, it is still unknown how a massive protostar can
continue to accrete despite the emmense outward radiation pressure it
exerts on its surroundings \citep[for reviews of massive star
  formation see e.g.][]{Beuther07,Hoare07}.

It has been suggested that massive stars may form through a mechanism
similar to that of lower mass stars, whereby matter is accreted from a
circumstellar disk, with the outward radiative force escaping via the
poles and driving a fast bipolar wind.  Several numerical studies have
succeeded in creating stars with masses $\ga$30\msun this way
\citep[e.g.][]{Y-S02,Krumholz09}, while it has been shown that the
fast bipolar winds should strongly influence the infalling material
from the envelope \citep{Parkin09}. The observational evidence to
support this scenario however is limited and circumstantial.  The CO
bandhead emission at 2.3\microns\ has been modelled as arising in a
circumstellar disk \citep[e.g.][]{Chandler93,Chandler95,B-T04,Blum04};
large-scale bi-polar outflows have been observed to originate from
heavily embedded objects \citep[e.g.][]{Beuther02}; and some authors
have found evidence of accretion disks in individual massive
protostars \citep{Patel05,Beltran06,Torrelles07}.




\begin{figure*}
  \includegraphics[width=17cm,bb=30 40 700 565]{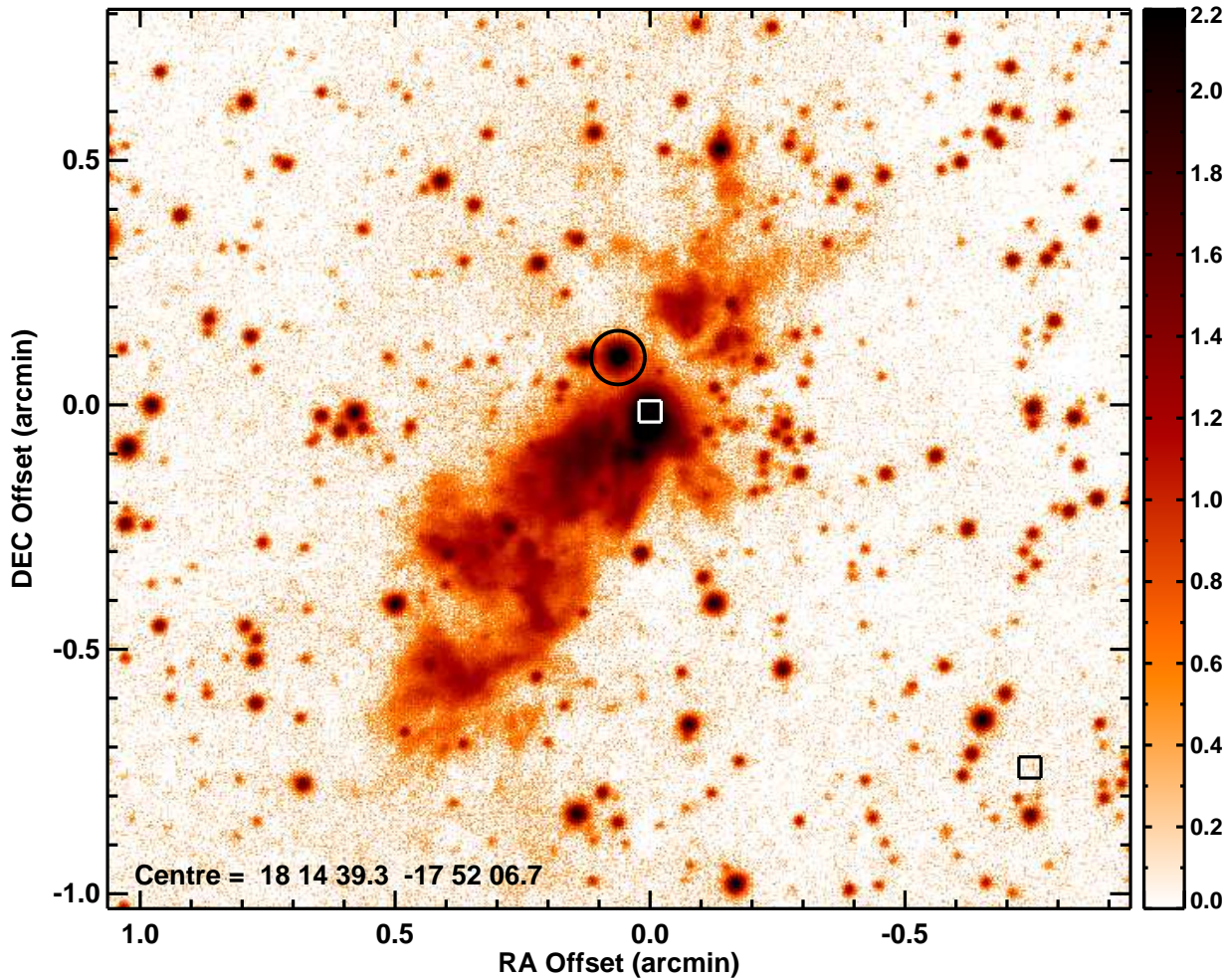}
  \caption{Wide-field $K$-band image of W33A, taken from the UKIDSS
    Galactic Plane Survey \citep{Lucas08}. The image is
    logarithmically scaled in units of sigma above the background. The
    NIFS field-of-view is indicated by the white box. Our sky field is
    indicated by the black box in the lower-right of the image. The
    natural guide star, used in the tip-tilt correction stage of the
    adaptive optics system, is indicated by the black circle. }
  \label{fig:wfim}
\end{figure*}

Massive Young Stellar Objects (MYSOs) are thought to represent the
phase in a massive star's formation immediately before the star
reaches the main-sequence (MS) but whilst it is still accreting, and
as such are important objects with which to study the formation of
massive stars.  W33A is a well-known MYSO. It lies at a distance of
$\sim$3.8kpc and has a source luminosity of $\sim$10$^5$\lsun
\citep{Faundez04}.  It is deeply embedded \citep[e.g.][]{Gibb00}, but
is visible in the $K$~band despite this. The morphology of the
object's large-scale nebula suggests a bipolar outflow, but as yet no
convincing evidence exists for an accretion disk or a bipolar wind
close to the central star. The object's weak radio emission has only
been marginally resolved at radio wavelengths
\citep{R-H96,vdTak05}. The close environment of W33A has been probed
using mid-IR interferometry \citep[][de Wit et al, 2009,
  submitted]{deWit07}, with the visibilities well fitted by a model in
which the emission arises in the warm dusty envelope close to the
surface of the cavity created by a bipolar outflow.

To provide a detailed study of W33A's inner nebula, and ultimately to
investigate the physical processes occuring in the formation of the
central star, in this paper we present high spatial resolution
integral field spectroscopy of W33A. These data allow us to study the
spatial variations in emission from the ionised outflow, the dense,
hot molecular material close to the star, and the cooler molecular
material in the object's inner envelope and outflow.

We begin in Sect.\ \ref{sec:obs} with a description of the
observations and data reduction steps. We present the results of the
data in Sect.\ \ref{sec:res}, and provide a discussion in
Sect.\ \ref{sec:disc}. We conclude in Sect. \ref{sec:conc}.

\section{Observations \& data reduction} \label{sec:obs}

\subsection{Observations}
In Fig.\ \ref{fig:wfim} we show the wide-field $K$-band image of the
W33A region taken from the UKIDSS Galactic Plane Survey
\citep{Lucas08}. The image clearly shows what appears to be a
large-scale outflow extending $\sim$0.8\arcmin\ towards the south-east
at a position-angle (PA) measured to be (135$\pm$5)\degr. Diffuse
emission is also seen to the north-west of the flux peak, which may be
the opposite lobe of the outflow. A velocity resolved $^{12}$CO map of
this region conclusively shows that the material is outflowing from
the central source (see Fig.\ 2 of de Wit et al., submitted). 

Our data were taken using the Gemini Near-infrared Integral Field
Spectrograph \citep[NIFS,][]{McGregor03} on the nights of 16 April and
25 May 2008. The observations were pointed such that the bright flux
centre was at the northern edge of the instrument field-of-view, in
order to obtain data on the base of the outflow as well as the central
source. The instrument applies the image-slicing technique to separate
the 3\arcsec$\times$3\arcsec\ field-of-view into 29 slices, which are
then aligned on the detector and passed through a diffraction
grating. We employed the K grating in combination with the HK filter
and central wavelength set at 2.2\microns. This setup achieved a
spectral resolution of $R \approx 5300$ over the wavelength range
2.0-2.4\microns\ across the full NIFS field-of-view.

Our observations made use of the Gemini ALTAIR laser-guide star (LGS)
adaptive optics system. The system uses a natural guide star for
tip-tilt correction, and the laser guide star for wavefront
sensing. For tip-tilt correction we used the star 2UCAC-25155527,
which has $m_{\rm UC} = 14.6$ and is located 6.9\arcsec\ from the
centre of our field-of-view (see Fig.\ \ref{fig:wfim}). 

The observational strategy involved observing the target and
neighbouring sky in an ABBA pattern, the location of the sky field is
shown in \fig{fig:wfim}. The integration time per individual
observation was 180s, with each integration read-out in the {\sc
  faint\_mode} setting of 16 non-destructive reads. A total of 7
separate integrations were used, giving a total integration time of
$\sim$20mins. In addition to the science target, we observed the A0~V
double HR6798 to characterize the telluric absorption. The GCAL
calibration unit was used to take continuum lamp observations plus
associated dark frames for flat-fielding purposes. Argon-Xenon arc
exposures were obtained for wavelength calibration, as well as ronchi
flats which are used to characterize the spatial distortion of NIFS
data.

\subsection{Data reduction}

The preliminary stages of data-reduction were done using the Gemini
instrument-specific pipeline which runs on the {\sc iraf}
platform. Science observations had their associated sky frames
subtracted to remove the dark current and sky emission lines. The data
were then divided through by the normalized flat-field to correct for
pixel-to-pixel variations in sensitivity, and any bad pixels flagged
in the earlier reduction steps were interpolated over. The arc and
ronchi flat frames were used to derive the transformation matrix of
the 2-dimensional `sliced' image onto a 3-dimensional ($x,y,\lambda$)
datacube. The application of this transformation rectifies the data
both spatially and spectrally, and it is at this step that the
wavelength calibration is applied. 

Removal of telluric absorption and combination of the individual
datacubes was done in IDL, using custom-written routines and the GSFC
IDL library\footnote{\tt http://idlastro.gsfc.nasa.gov}. Correction
for the \brg\ absorption in the telluric standard was done by dividing
through by a synthetic AO~V template spectrum, interactively altering
the centroid and absorption strength of the \brg\ feature in the
template until a satisfactory match to the standard star was found. At
each spatial pixel of the science datacube, the spectrum was extracted
and divided by the telluric standard. If the spectrum had sufficient
signal-to-noise (i.e.\ over $\sim$100 counts in the continuum) the two
spectra were first cross-correlated to achieve optimal telluric
correction, particularly important in the region of the OH absorption
lines at 2.0-2.1\microns\ where sub-pixel misalignments can introduce
artifacts into the spectrum.

The individual science datacubes of repeat observations were spatially
aligned by first taking the median image of each cube across all
spectral pixels, then cross-correlating these median images with one
another. We then took the sum of the spectra at each spatial pixel,
using the median of these spectra as a template from which to reject
any further bad pixels and cosmic-ray hits.

\begin{figure}
  \includegraphics[width=8.5cm,bb=40 10 638 595]{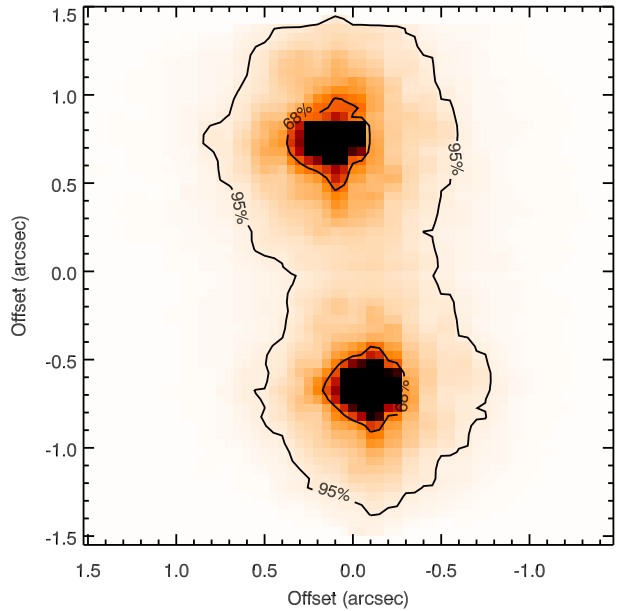}
  \caption{Image of the A0~{\sc v} double HR6798A+B, illustrating the
    performance of the ALTAIR-LGS adaptive optics system during the
    observations. Contours show the encircled energy at 68\% and 95\%
    of the total flux. The FWHM was measured as 0.13\arcsec. }
  \label{fig:psf}
\end{figure}

\subsection{Adaptive optics performance}
In order to illustrate the spatial resolution of the data and the PSF
correction that was achieved through the ALTAIR adaptive optics
system, in Fig.\ \ref{fig:psf} we show the wavelength-integrated image
of the telluric standard stars HR6798A+B. To better demonstrate the shape
and size of the corrected PSF, the image was first split into two,
with each sub-image containing one star. The sub-images were
normalized such that the corrected flux from each star was the same,
then the sub-images were recombined. On the image we have drawn
contours indicating 67\% and 95\% encircled energy. While the star to
the north appears slightly elliptical, the southern star is circular
to within the errors. For each star we find 67\% of the PSF flux is
contained within a radius of $\approx$0.15\arcsec\ of the flux
centre. The full-width half-maximum (FWHM) of each star was measured
to be 0.13\arcsec.

\subsection{Variations in spectral resolution and wavelength calibration}
The accuracy of the wavelength calibration and the consistency of
spectral resolution was determined from analysis of datacubes
constructed from the arc lamp observations. For each arc line that was
used to determine the wavelength solution we fitted gaussian line
profiles, and measured the line-width and velocity offset. We found
that the absolute wavelength calibration had a root-mean-square
standard deviation of $\pm$4\kms\ across the field-of-view, with no
discernable trends with spatial position. The spectral resolution was
similarly found to be $\Delta v = 55 \pm 4$\kms\ across the field,
again having no trend with spatial position, other than that it was
systematically poorer by $\sim$6\kms\ in the south-east
0.5\arcsec$\times$0.5\arcsec region. As very little data was used from
this region, we did not correct for this effect.

\section{Results} \label{sec:res}

\begin{figure}
  \includegraphics[width=8.5cm,bb=-20 0 660 605]{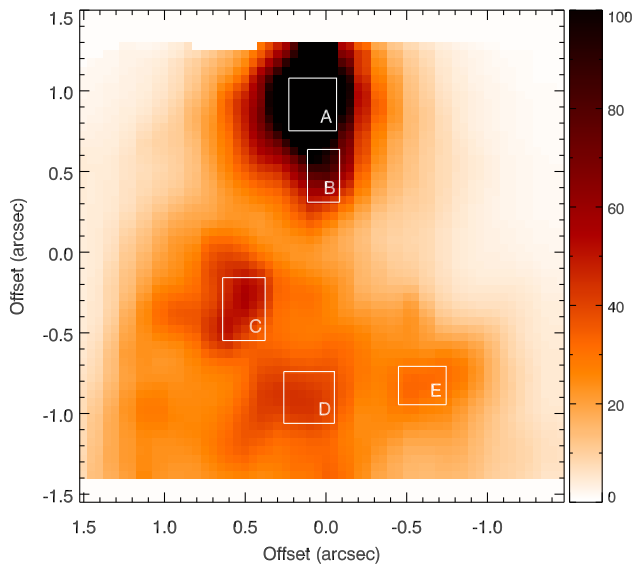}
  \caption{NIFS K-band image of W33A when integrated over all spectral
    channels, with North up and East to the left. The image is scaled
    in units of sigma above the background, estimated from the NW
    corner of the field. The location of maximum flux is labelled `A',
    and the other four bright knots of emission are labelled
    `B'-`E'. }
  \label{fig:imcont}
\end{figure}

\begin{figure}
  \includegraphics[width=8.5cm]{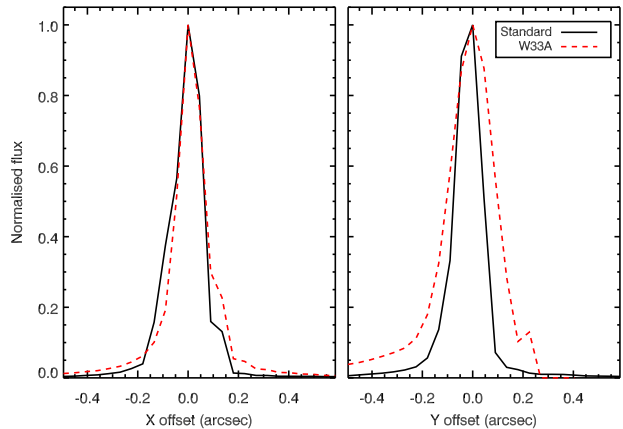}
  \caption{Spatial profile of the flux centroid of W33A when compared
    to the average profile of the two standard stars. While the
    profiles in the $x$-direction are similar, the $y$-profile of W33A
    is noticably extended in comparison to the standard. }
  \label{fig:psfprof}
\end{figure}

\subsection{Nebular morphology}
In Fig.\ \ref{fig:imcont} we show the total two-dimensional image when
the datacube is summed across all wavelength channels. The central
source is not a point-source, and instead is noticeably extended in
the north-south direction (see Fig.\ \ref{fig:psfprof}). In
particular, we see a bright central peak (labelled `A' in
Fig.\ \ref{fig:imcont}) and a `finger' pointing southward (labelled as
'B'). The `L'-shaped inner nebula, which can be marginally resolved in
the UKIDSS $K$-band image and which forms the base of the outflow, is
resolved in our data into three bright knots of emission, which we
label `C', `D' and `E'. As we show later, the emission from these
knots is predominantly scattered light.


\begin{figure*}
  \includegraphics[width=17cm,bb=0 0 1275 793]{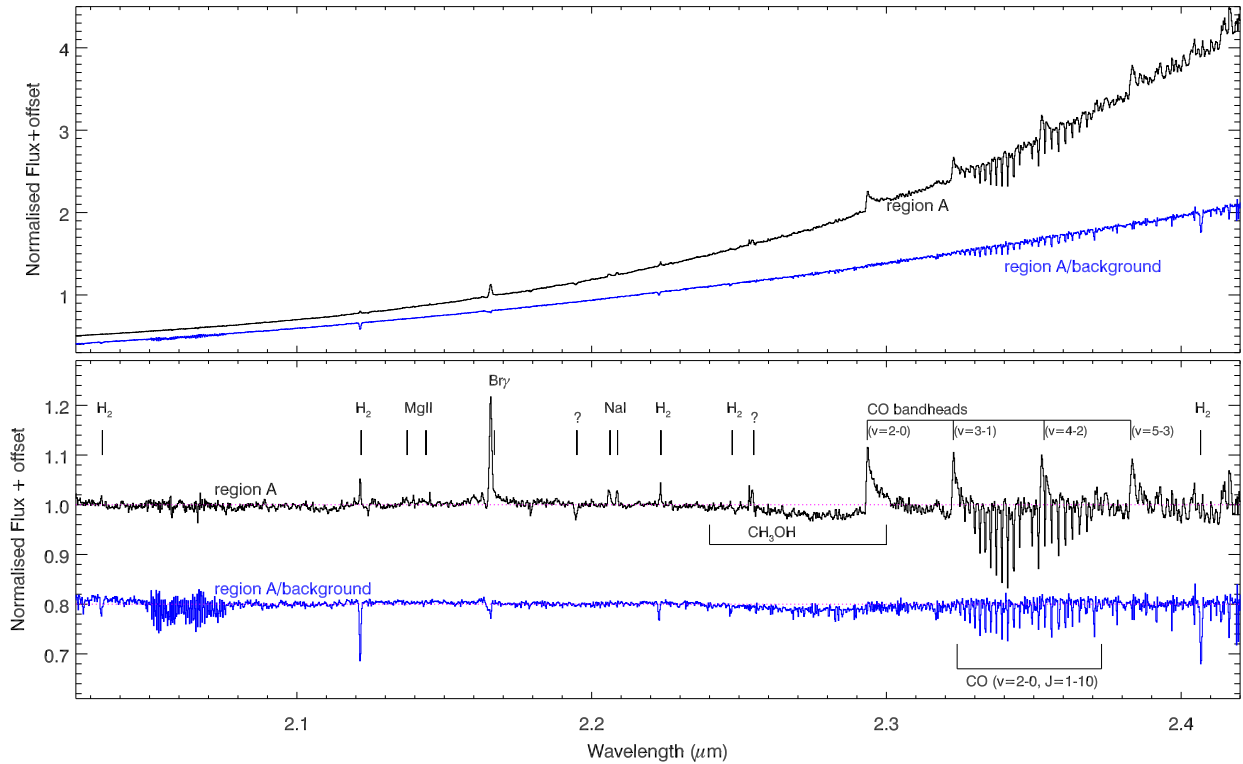}
  \caption{Spectrum of region A, and the ratio of region A's spectrum
    to that of the average background spectrum. The top panel shows
    the spectra before continuum fitting, to illustrate the high
    degree of reddening. The bottom panel shows the spectra after
    dividing through by fits to the continua, to highlight the
    discrete spectral features.}
  \label{fig:totsp}
\end{figure*}

\subsection{Spatial variations of spectral
  properties} \label{sec:specprop} 

The unique feature of IFU data is that it is possible to chart the
variations of the observed spectral properties across the field of
view. We begin in this section with a description of the principle
spectral features, followed by an analysis of how these features vary
with spatial position. 

\begin{table}
  \caption{Equivalent widths of the atomic emission lines. The
    uncertainty on EW was measured from fluctuations in the
    continuum. }
  \label{tab:ew}
  \centering
    \begin{tabular}{lcc}
      \hline \hline
      Line & $\lambda_{\rm vac}$ & EW (\AA) \\
      \hline
      Br\,$\gamma$ & 2.1662 & -5.00$\pm$0.17 \\
      Mg\,{\sc ii} & 2.1374 & -0.43$\pm$0.17 \\
      Mg\,{\sc ii} & 2.1438 & -0.24$\pm$0.17 \\
      Na\,{\sc i} & 2.2062 & -0.50$\pm$0.17 \\
      Na\,{\sc i} & 2.2090 & -0.37$\pm$0.17 \\
      \hline    
    \end{tabular}
\end{table}

In Fig.\ \ref{fig:totsp} we plot the spectrum of the bright central
source extracted from region A (see Fig.\ \ref{fig:imcont}). The
steeply-rising continuum indicates that the central source is heavily
reddened. In order to highlight the spectral features, in the bottom
panel of the figure we show the spectrum when normalized by a
4th-degree polynomial fit to the continuum (labelled as `region A' in
the Figure). In this normalised spectrum we see emission features of
\brg, H$_2$, \nai\ and the CO 2-0, 3-1, 4-2 and 5-3 bandheads. Also
seen are narrow absorption lines in the region of the CO bandheads,
attributable to low J transitions of the CO v=2-0 state, as well as a
broad absorption feature around 2.28\microns\ which we attribute to
\choh\ ice \citep[also observed in this object's spectrum
  by][]{Taban03}. We do not confirm the presence of \hei\ emission
detected by \citet{Taban03}, while there is a suggestion of emission
from the \mgii\ doublet at 2.14\microns. Table \ref{tab:ew} shows the
measured equivalent widths of the atomic emission lines as measured
from the region A aperture.

\subsubsection{Spectrum of the central source versus surrounding nebula}
To analyse spectral differences between region A and the rest of the
field, we have created an average `background' spectrum by integrating
over all spatial pixels in the observed field {\it except} those in
region A. In the top panel of Fig.\ \ref{fig:totsp} we have also
plotted the ratio of the region A spectrum to this background
spectrum, labelled `region A/background'. The continuum of this ratio
spectrum is shallower than that of the central source, which is to be
expected if the emission in the southern nebula is predominantly
scattered light. 

In the bottom panel of \fig{fig:totsp} we plot the ratio of the
spectra when normalized by a continuum fit. Here the emission due to
\brg, \nai, \mgii\ and CO disappears, suggesting that the emission
morphologies of these features follow that of the continuum,
i.e.\ they are also scattered. The H$_2$ lines go into `absorption'
indicating that this emission originates mainly in the surrounding
nebula rather than on the bright source at region A. The \choh\ and CO
absorption are still seen in the residual spectrum, implying that
there is an increased level of opacity due to these features along the
line-of-sight to region A compared with the surrounding nebula.


\begin{figure}
  \includegraphics[width=8.5cm,bb=20 20 788 723]{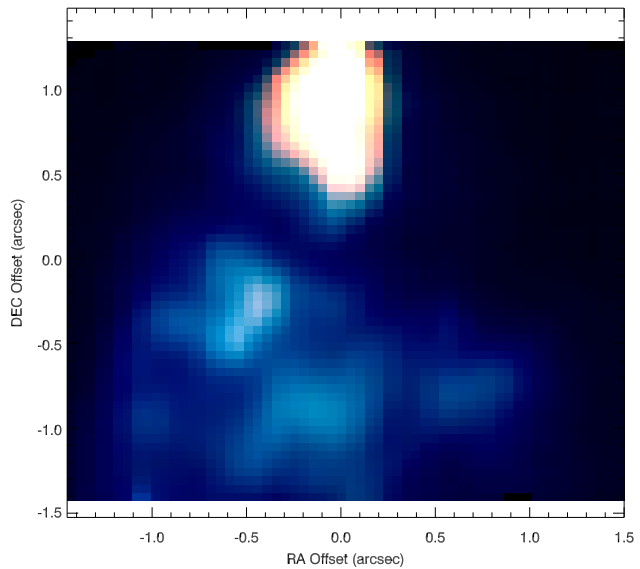}
  \caption{Three-colour image of the continuum emission. Each RGB
    channel is scaled linearly, while the dynamic ranges were chosen
    arbitrarily to best illustrate the contrast between region A and
    the southern nebula.}
  \label{fig:3col}
\end{figure}

\begin{figure*}
  \includegraphics[width=8.5cm,bb=-20 0 708 643]{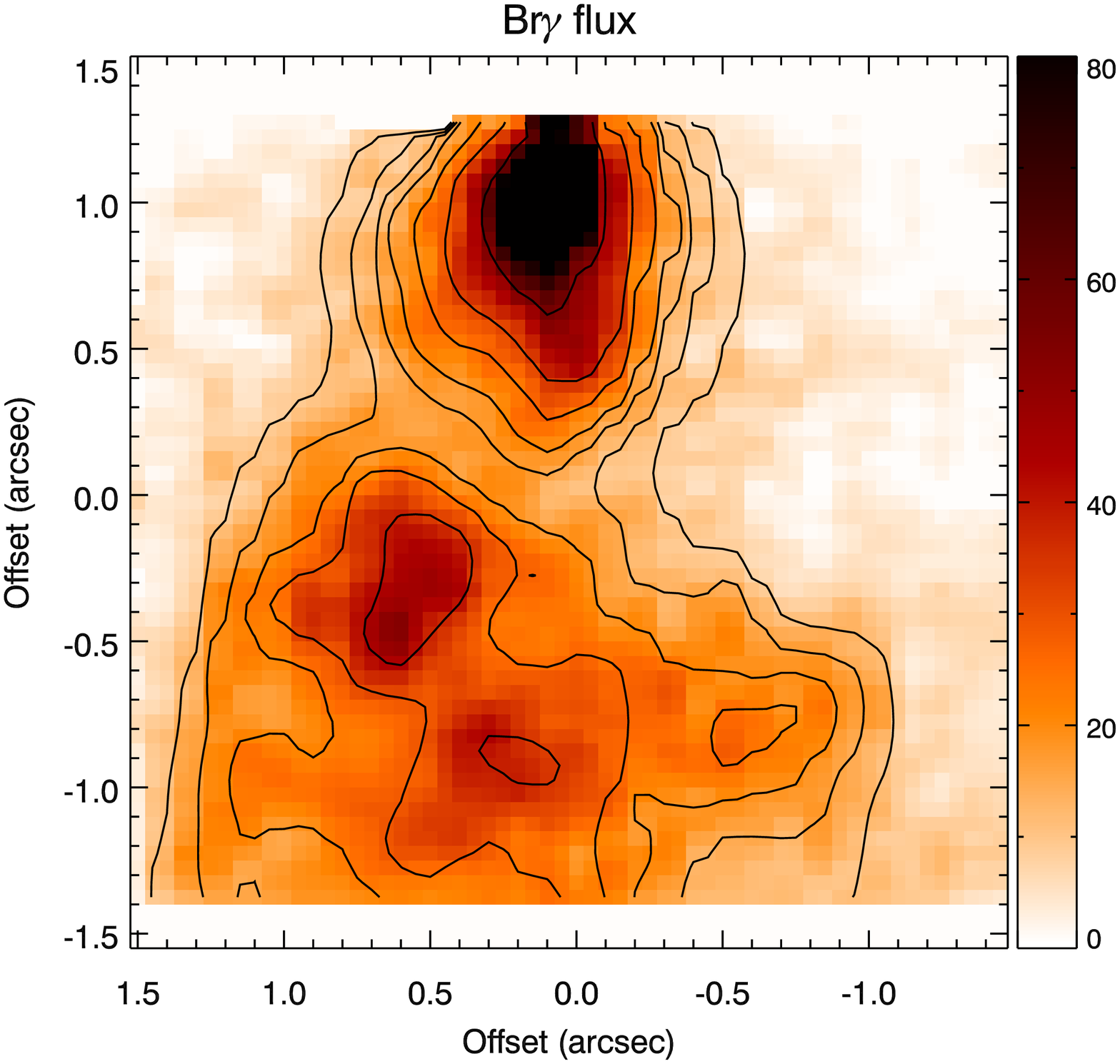}
  \includegraphics[width=8.5cm,bb=-20 0 708 643]{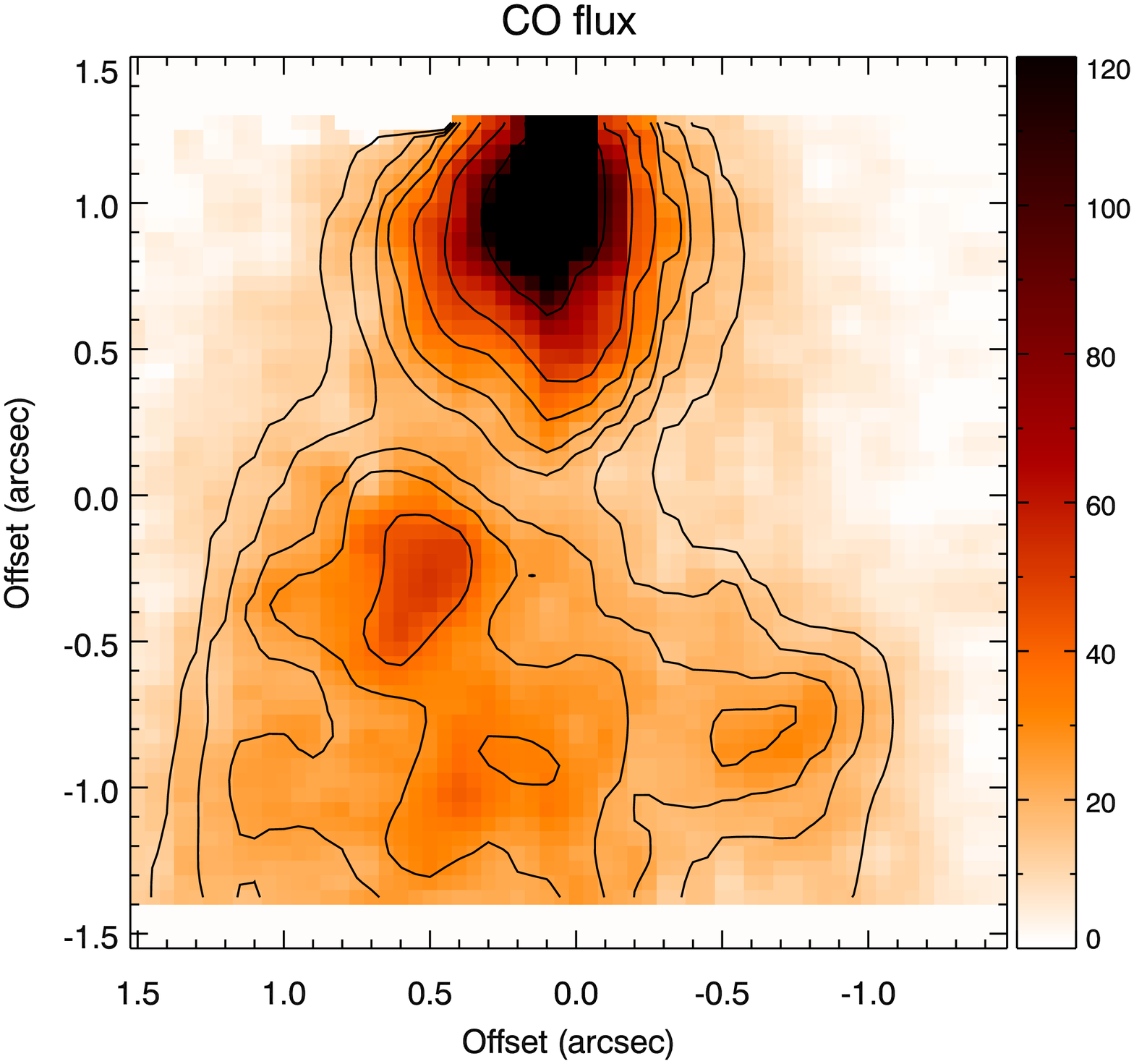}
  \caption{{\bf Left:} Image of the continuum-subtracted line flux of
    \brg. The images are scaled in units of sigma above the
    background, measured from the north-west corner of the image. The
    overplotted contours show the continuum emission and are drawn at
    20$\sigma$, 30$\sigma$, 40$\sigma$, 50$\sigma$, 70$\sigma$,
    100$\sigma$ and 200$\sigma$ above the background. {\bf Right:} Same
    as the left-hand figure but for the first CO bandhead at
    2.293\microns. }
  \label{fig:brgim}
\end{figure*}

\subsubsection{Variations in continuum emission}
In the previous section we showed that the slope of the $K$-band
continuum emission is not uniform across the field. To further
illustrate the spatial variations in continuum emission we created
narrow band continuum images at three featureless regions of the
spectrum: 2.083\microns, 2.185\microns, and 2.273\microns, each with a
bandwidth of $\pm$0.003\microns. These images were then combined into
an RGB image, with the shortest wavelength image assigned to the blue
channel and the longest to the red. The resulting image is shown in
Fig.\ \ref{fig:3col}.

In the image, the southern nebula is seen predominantly in `blue'
light (i.e.\ the shortest wavelength), while the central source
appears red. This behaviour can be understood as a combination of
scattering and extinction. The central source at region A is heavily
absorbed, making the direct light from region A appear red. Meanwhile,
the emission to the south is mostly light from region A which has been
scattered, and so appears blue.

\subsubsection{The \brg\ and CO emission} \label{sec:brgco}
In this section we map the spatial morphology of the emission due to
\brg\ and CO (the CO absorption is studied in
Sect.\ \ref{sec:coabs}). First we subtract the continuum at each
spatial pixel of the datacube by fitting a 4th-degree polynomial to
featureless regions of the continuum. We then integrate the emission
across the spectral feature in question. For the CO emission, we took
only the bright blue edge of the first bandhead (2.293-2.300\microns)
-- this feature contains the majority of the flux, and is
uncontaminated by the low-J absorption lines.

The continuum-subtracted emission images of \brg\ and CO are shown in
Fig.\ \ref{fig:brgim}. Both images display morphology similar to that
of the continuum image (traced by the contours), with each of the five
emission knots clearly visible. It is unlikely that the extended
\brg\ and CO emission is formed `in-situ' in the southern parts of the
nebula, as the two spectral features require different temperatures
and densities. It is far more likely that in the southern nebula, as
with the continuum emission, we are seeing scattered light which
originates within a few AU of the central source.

\begin{figure}
  \includegraphics[width=8.5cm,bb=-20 0 708 643]{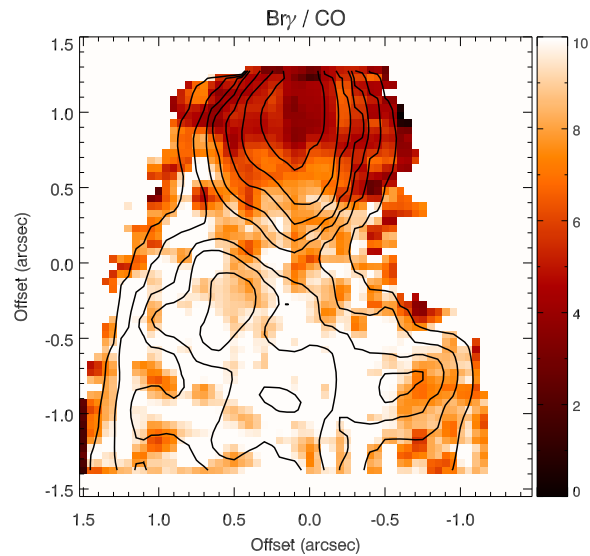}
  \caption{The ratio image of the continuum-subtracted \brg\ and CO
    emission. Contours are the same as those shown in \fig{fig:brgim}.}
  \label{fig:brg-co}
\end{figure}

In Fig.\ \ref{fig:brg-co} we further explore the scattered light
explanation for the \brg\ and CO emission by taking the ratio of the
two continuum-subtracted images. If the emission in the southern
nebula attributable to these transitions is due to scattering by
material in the outflow, we should in principle see very little
morphological difference between the emission
morphologies. \Fig{fig:brg-co} shows that no features corresponding to
the southern nebula are seen in the ratio image. Instead, we see a
slight increase in the ratio of \brg\ to CO in the southern regions
compared that around region A. This can be understood as a combination
of the wavelength dependence of scattering, which boosts the \brg/CO
ratio in the south, and extinction by dense material around the
central source, which attenuates \brg\ with respect to CO around
region A.

\begin{figure*}
  \includegraphics[width=17cm]{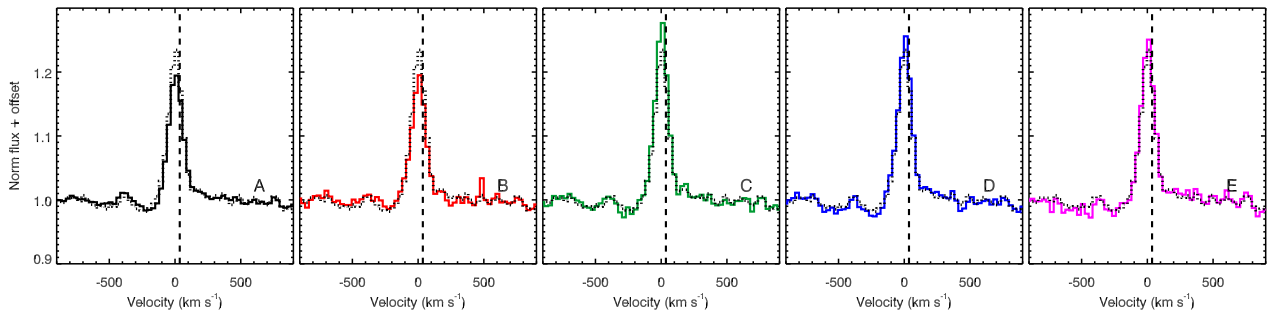}
  \caption{The variations of the \brg\ spectral feature across the
    five bright knots of emission identified in
    Fig.\ \ref{fig:imcont}. The mean of the five spectra is
    overplotted as a dotted line. The dashed line indicates W33A's
    systemic velocity.}
  \label{fig:brgabc}
\end{figure*}
\begin{figure*}
  \includegraphics[width=17cm]{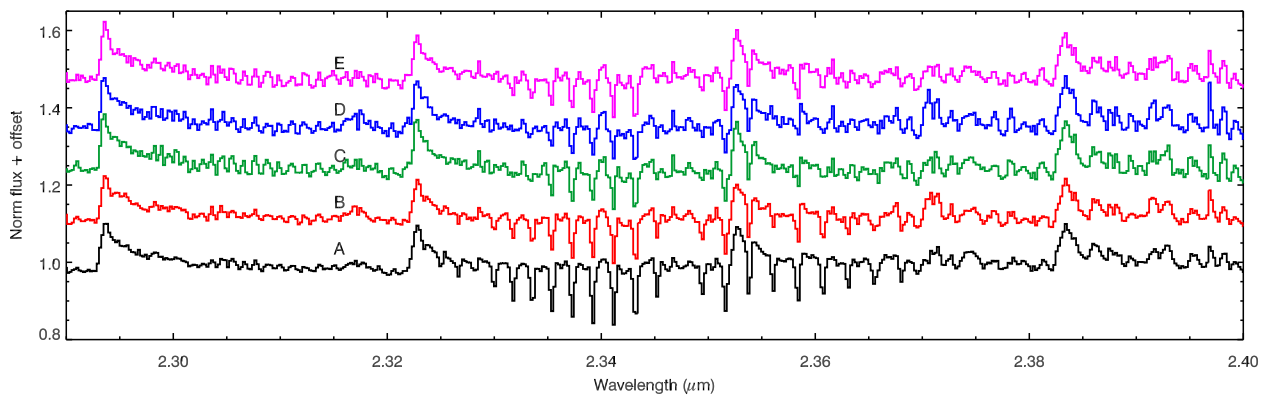}
  \caption{The variations of the CO feature across the five bright
    knots of emission identified in Fig.\ \ref{fig:imcont}.}
  \label{fig:coabc}
\end{figure*}
\begin{figure*}
  \includegraphics[width=17cm]{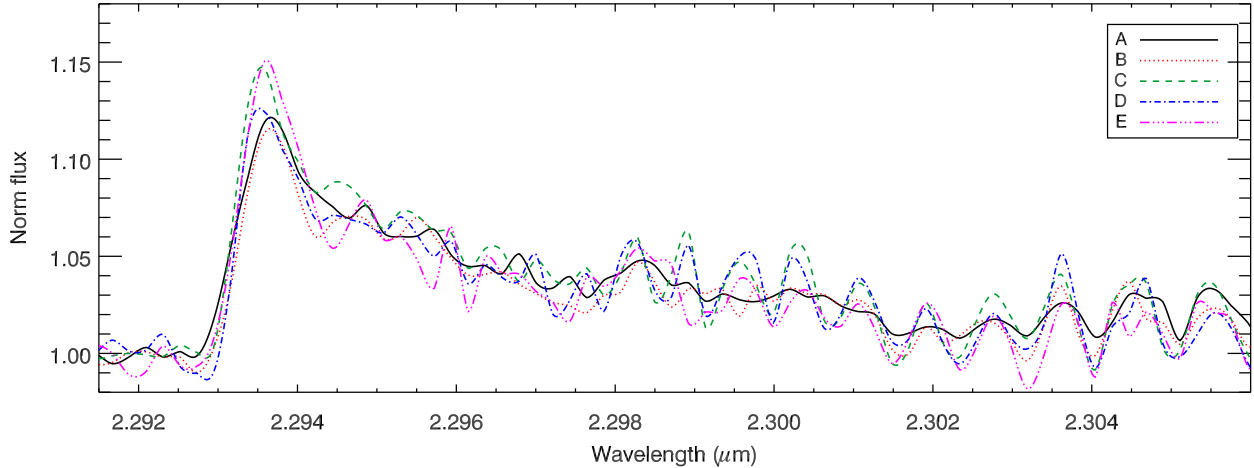}
  \caption{Spectra of the five regions identified in
    Fig.\ \ref{fig:imcont} around the CO v=2-0 bandhead. For clarity,
    spectra have been interpolated onto a finer grid. It is clear from
  this figure that the line profiles of C, D and E are narrower and
  more pronounced than those of A and B.}
  \label{fig:coblue}
\end{figure*}

\subsubsection{Variations in line-profile with viewing-angle}
As we have shown in previous sections, the line-emission observed in
the southern nebula is most likely scattered light which orginates at
the flux centre (i.e. region A). We can then use this scattered light
to study the emission from the central source as viewed by the
southern nebula, when compared to viewing the emission directly from
region A. In effect, this allows us to study the line-profiles as a
function of viewing angle. This is a technique we have previously used
to study the wind geometry of the evolved massive star IRC~+10420
\citep{IRCpaper}.

We explore the spatial variations of \brg\ and CO in
Figs.\ \ref{fig:brgabc} and \ref{fig:coabc}, where we plot the
integrated spectra of the five bright knots of emission identified in
Fig.\ \ref{fig:imcont} as A-E. In respect to the \brg\ emission, there
is very little variation across the five regions. The regions show
evidence for P~Cygni absorption, and perhaps an extended red wing. The
equivalent widths (EWs) of the profiles in the southern nebula are
slightly larger than those of regions A and B. As the \brg\ emission
in the southern nebula is most likely scattered light, these small
changes in EW are possibly due to anisotropy in the ionized
circumstellar gas.

The CO emission shows clear variations across the observed field.  To
highlight this in Fig.\ \ref{fig:coblue} we plot the spectra in the
vicinity of the v=2-0 bandhead. The blue edge appears shallower in
regions A and B, while the edge is much steeper in the three southern
regions. Also, the transitions between 2.30-2.305\microns\ appear to be
narrower and more pronounced in the southern knots (C, D and E)
than in those close to region A. This suggests that the CO emission,
when viewed directly through the bright central source at region A,
has a broadened velocity profile with respect to the profile reflected
from the southern nebula. As the southern nebula's viewing angle to
the central source is different from our direct line of sight to the
same region, this suggests that the CO profile's broadening is a
function of the viewing angle. Such behaviour is typically attributed
to the emission forming in a disk \citep{B-T04,Blum04}. As the
material in the southern nebula `sees' less rotational broadening than
our direct line-of-sight, the orientation of the disk is consistent
with being aligned perpendicularly to the larger-scale outflow.

\begin{figure*}
  \includegraphics[height=8.3cm]{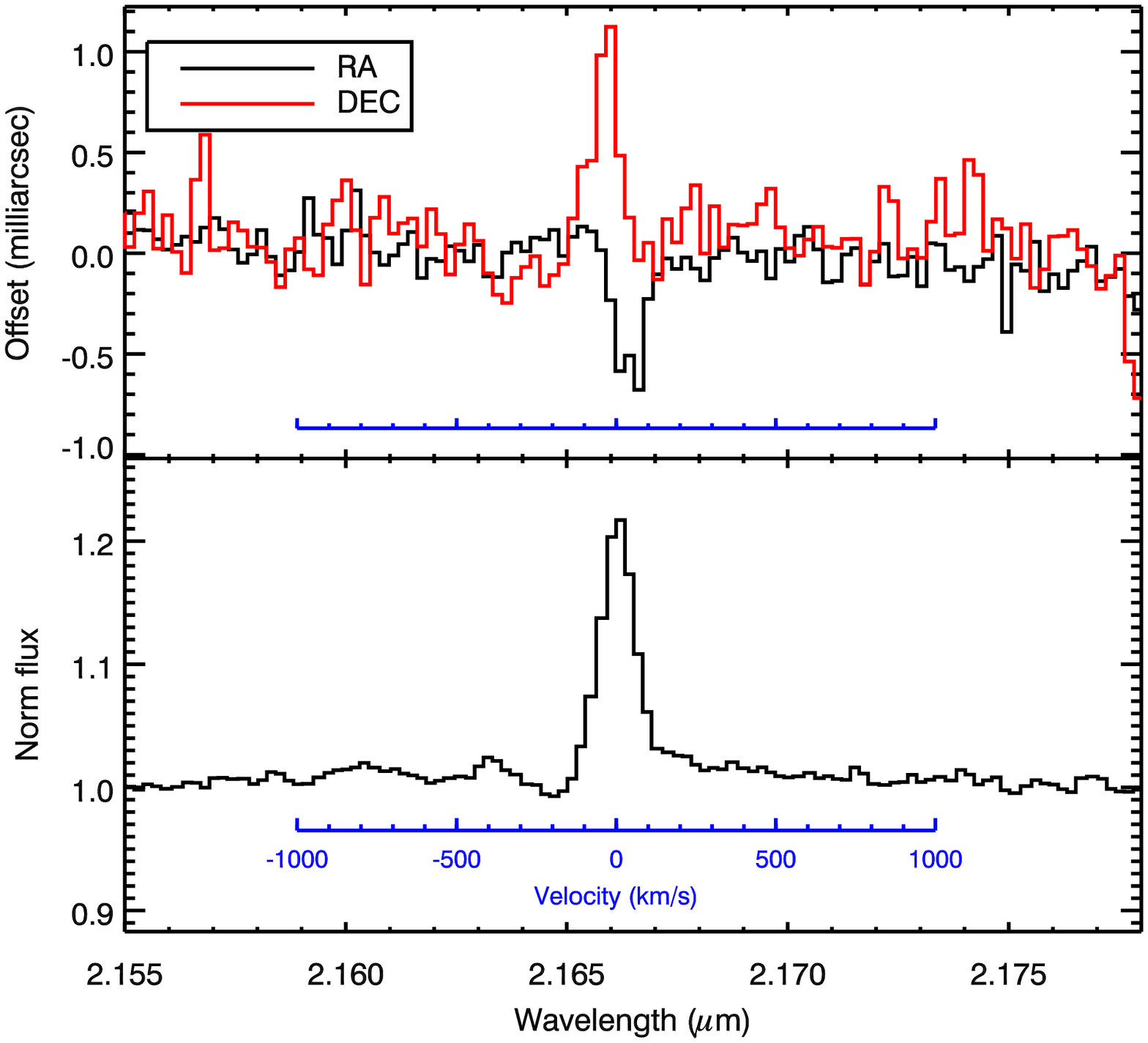}
  \includegraphics[height=8.3cm]{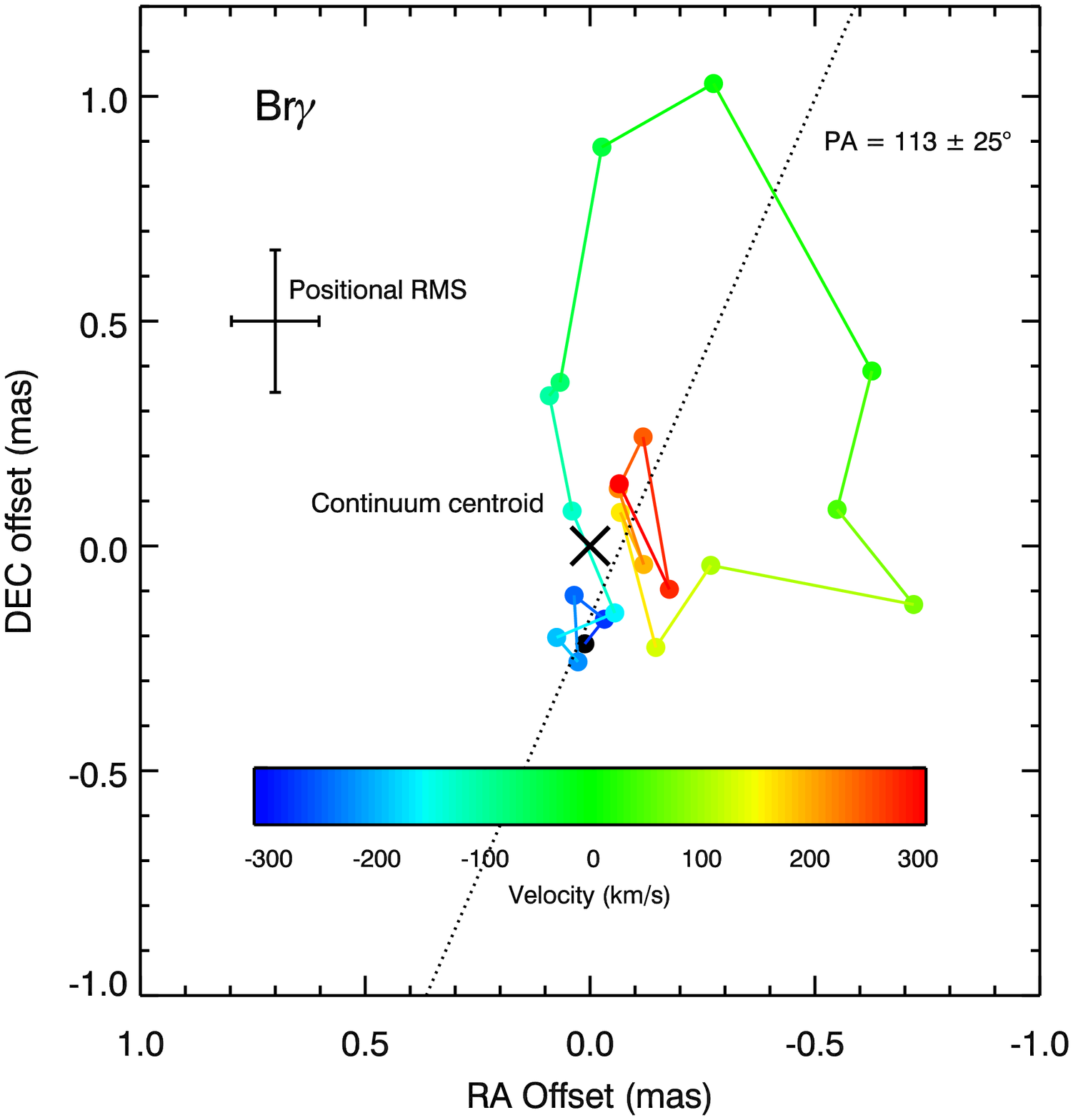}
  \caption{Spectro-astrometry across the \brg\ emission line. ({\bf
      Left}): the bottom panel shows the intensity spectrum of region
    A, while the top panel shows the astrometric position of the
    centroid flux peak as a function of wavelength. The red line,
    labelled `DEC', shows the centroid position in the $y$-direction,
    while the black line ('RA') shows the position in the
    $x$-direction. ({\bf Right}): the same data as in the left-hand
    panel, but represented in $x-y$ space. The symbols are
    colour-coded to show the velocity at each point. The dotted line
    shows the position-angle of the red and blue high-velocity
    components. }
  \label{fig:specast}
\end{figure*}

\subsubsection{Spectro-astrometry of \brg} \label{sec:specast}
Here we investigate the small-scale geometry of the \brg\ emission by
performing a 3-D spectro-astrometric analysis of the \brg\ line. At
each wavelength channel of the datacube we extracted a 2-D image and
determined the astrometric centroid of the emission at region A. This
method can give spatial information on extremely small scales -- the
astrometric accuracy on the centroid is roughly equivalent to $\Delta
\sim {\rm FWHM}/{\rm SNR}$. So, for the spatial resolution of our
observations (0.13\arcsec) and a SNR at region~A of several hundred,
in principle we can expect to achieve accuracies of order of a tenth
of a milliarcsecond. In order to verify this we performed
spectro-astrometric tests on the data from the telluric standards. We
found that indeed the r.m.s. on the centroid position was of order
0.1mas. We tried different methods of centroiding, including
flux-weighted centroiding, finding the location at which the
derivatives go to zero and fitting 2-D gaussian profiles. All
techniques were found to give similar results within the errors. In
addition we observed no artifacts around the telluric absorption
lines, and no excursions across the telluric standard star's intrinsic
\brg\ absorption, as is to be expected (see Appendix~A).

In the left panel of \fig{fig:specast} we plot the position of the
centroid as a function of wavelength in the region of \brg. The
spectro-astrometric profiles have had any large-scale gradients
removed by fitting a 3rd-degree polynomial to the continuum. Firstly,
we note that the astrometric accuracy of the centroid, as judged by
the continuum regions, is comparable to that predicted (0.1-0.2mas),
corresponding to a physical scale of 0.4-0.8AU for a systemic distance
of 3.8kpc. Secondly, the centroid shows a clear shift in the region of
\brg. In the north-south direction (labelled `DEC' in the figure), the
centroid appears to shift south slightly in the region of the P~Cygni
absorption, before shifting northwards by $\sim$1\,mas. Similarly, in
the east-west direction (labelled `RA', where the positive direction
is eastward), there is a slight shift to the east coincident with the
P~Cygni absorption, followed by an excursion to the west of
$\sim$0.5mas.

This behaviour is illustrated again in the right panel of
\fig{fig:specast}. This time, the centroid is plotted in RA-DEC space
with colour-coded symbols to indicate the velocity of each point. The
location of the continuum centroid is illustrated by the cross. The
plot shows that the emission with the largest red-shifted velocities
($\sim$400\kms) is located to the north-west of the continuum
location, while the emission blue-shifted by a similar amount is
located to the south-east of the continuum, indicating expansion. From
analysis of those points with speeds between 300-400\kms\ we find that
the statistical significance of the red- and blue-shifted points being
spatially separated from the continuum is 2.8$\sigma$ and 4.1$\sigma$
respectively. In addition to this behaviour there is a large excursion
across the centre of the line with $|v| < 100$\kms.

The observed behaviour can be explained if we make the assumption that
the \brg\ emission line is made up of two or more components. The
broad high-velocity emission originates in an axisymmetric structure,
most likely a bi-polar jet, while the narrow low velocity emission is
formed in an extended structure such as a small, dense
\hii-region. The nature of this narrow emission is difficult to
disentangle, but it is possible that it is formed in an aspherical
clumpy medium with much velocity-dependent extinction and
self-absorption. This is similar to the picture deduced by
\citet{Drew93} and \citet{Bunn95} from velocity-resolved line-profile
ratios. 

In addition to the detection of an outflow, the spectro-astrometric
data also indicate that the bright flux maximum observed in the
integrated image is indeed the location of the central star. 

For the broad-line component of the emission we can measure the PA of
this structure by taking the data at $300<|v/{\rm km\,s^{-1}}|<400$
and taking the mean positions of the red- and blue-shifted
components. We measure the PA of this structure to be
113$\pm$25\degr\ (see right panel of \fig{fig:specast}), and so is
aligned with the large-scale outflow seen in \fig{fig:wfim}
(PA=135$\pm$5\degr). From the alignment of these angles it is likely
that in this small-scale structure we are seeing the base of a
bi-polar ionized wind or jet. In addition, the blue-shifted portion of
the \brg\ emission is oriented towards the south-east, consistent with
the hypothesis that the outflow seen in the wide-field image is the
lobe coming towards us, while the red-shifted lobe to the north-west
is largely obscured.

The outflow velocity, as measured from \fig{fig:specast}, appears to
be around $v_{\rm outflow} \cos i \sim 300$\kms. Assuming an
inclination of 60\degr\ (de~Wit et al., 2009 submitted), this gives an
outflow velocity measured from \brg\ of $\sim$600\kms. This is likely
an underestimate of the outflow's terminal velocity, as a large
fraction of the \brg\ emission may form in the accelarating part of
the wind. This value is consistent with MYSO jet velocities measured
by radio proper motion studies \citep{Marti98,Curiel06}.

\begin{figure}
  \includegraphics[width=8.5cm]{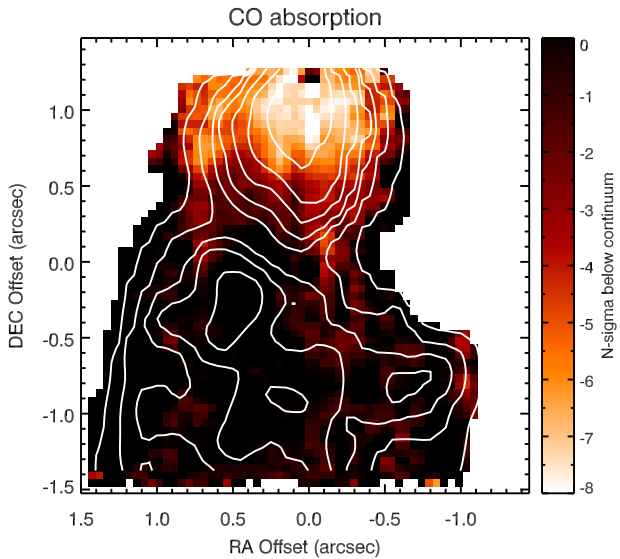}
  \caption{Image of the datacube integrated over the CO absorption
    lines. The image is scaled in units of sigma below the continuum
    flux level, where sigma is measured from the southern nebula.  }
  \label{fig:co_absint}
\end{figure}

\subsubsection{The CO absorption} \label{sec:coabs}
The low-J transistions of the CO v=2-0 state are observed in
absorption in W33A's spectrum (Sect.\ \ref{sec:specprop}). The
population of these states requires much cooler conditions
($\sim$30\,K) than that of the high-J states seen in emission (few
$\times$1000\,K) \citep{Chandler93,Chandler95,B-T04}. Therefore these
two distinct spectral features must arise in different regions -- the
CO bandhead emission likely originates in material close to the
central star; while the absorption lines are formed in the outer
envelope at several thousand AU (see later).

The right-hand panel of Fig.\ \ref{fig:brgabc} shows that the CO
absorption lines vary in strength between the bright central source
(A) and the southern regions (C, D and E). Specifically, the
absorption is stronger towards region A than the southern nebula. To
explore this behaviour further, sub-cubes were extracted from the
primary data-cube, each containing the full spatial information of one
of these absorption lines. We observed a behaviour which was similar
for each line and so all lines were combined to increase the
signal-to-noise.

In \fig{fig:co_absint} we show the image obtained when the datacube is
integrated over these features. The image shows that the CO absorption
is concentrated around region A in a band extending roughly east-west,
and is perpendicular to the large-scale outflow. Significantly less
absorption is seen in the southern nebula, indicating that the
material in the southern regions sees less obscuring material than our
direct line-of-sight to region A. This suggests that the envelope has
the morphology of a torus, with a physical radius of $\sim$2000\,AU,
and with an opening to the south through which the outflow escapes.


\begin{figure}
  \includegraphics[width=8.5cm,bb=10 0 680 515]{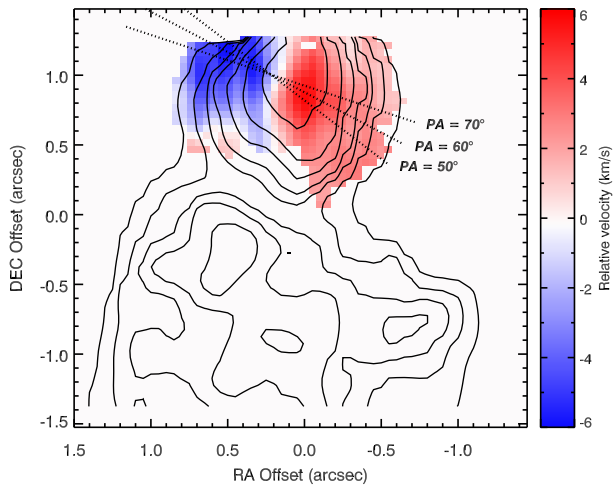}
  \caption{The velocity centroid of the CO absorption lines as a
    function of spatial position across the field. Velocities are
    indicated only for those absorption lines which have strengths
    greater than 5$\sigma$ below the continuum. The data has been
    smoothed with a boxcar filter of width 5 pixels (0.25\arcsec) to
    accentuate the regions of maximum and minimum velocity. Dotted
    lines indicate the rough position angle of the structure. }
  \label{fig:co_velogram}
\end{figure}

\Fig{fig:co_velogram} shows the measured velocity centroid of the CO
absorption at each spatial position across the extinction structure
seen in \fig{fig:co_absint}. The image has been smoothed by a boxcar
filter of width 5 pixels to highlight any velocity gradient. At
negative (blue-shifted) velocities, the opacity appears to be
concentrated in a region about 0.2-0.3\arcsec\ east of the centre of
the bright source (i.e.\ region A). At positive (red-shifted)
velocities however, there is a clear shift in the location of the
opacity centre to a region 0.1-0.2\arcsec\ {\it west} of region A. No
apparent trends with velocity are seen in the southern regions of the
nebula, though the data are very noisy and are not shown in the
figure. We determined the PA of rotation to be 60$\pm$10\degr\ by
fitting the blue and red flux peaks in \fig{fig:co_velogram}. This is
therefore consistent with being perpedicular to the larger scale
outflow of PA=135$\pm$5\degr. The PA of this feature is perpendicular
to the large scale outflow observed in \fig{fig:wfim}, as well as to
the small-scale outflow seen in the spectro-astrometric analysis of
the \brg\ emission. As with the broadening of the CO bandhead emission
described in the previous section, the observed behaviour is
indicative of an extended rotating structure seen edge-on. The
kinematics of this feature are explored further in
Sect.\ \ref{sec:disc}.

\subsubsection{Molecular hydrogen}
\begin{figure}
  \includegraphics[width=8.5cm,bb=-20 0 708 643]{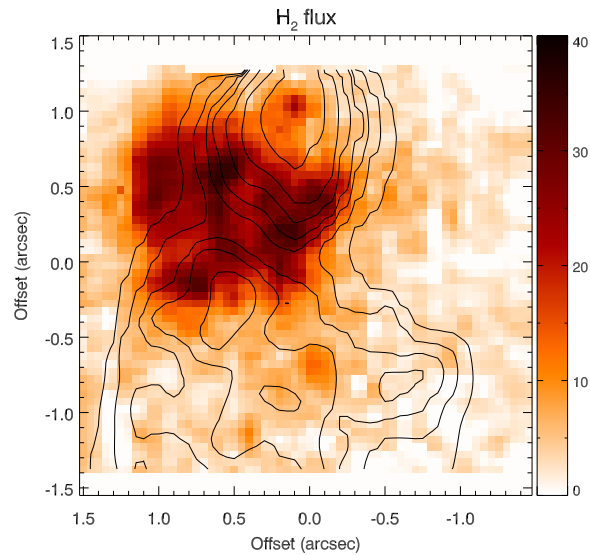}
  \caption{Same as Fig.\ \ref{fig:brgim} but for H$_2$
    2.12\microns. The point-like emission at the very centre of region
    A is possibly due to poor continuum subtraction. }
  \label{fig:h2im}
\end{figure}

The light of H$_2$ 2.12\microns\ shows a very different morphology to
that of the continuum and the \brg\ and CO emission (see
Fig.\ \ref{fig:h2im}). The emission is concentrated to the south-east
of region A, while very little H$_2$ emission is seen at region A
itself. The other H$_2$ lines observed in the spectrum display
similar behaviour, though have much poorer signal-to-noise and so are
not shown. The orientation of this emission, extending roughly
south-eastwards of region A, is approximately coincident with that of
the base of the large-scale outflow seen in the wide-field image of
\fig{fig:wfim}.

\section{Discussion -- CO kinematics, and the mass of the central object} \label{sec:disc}

The pieces of evidence presented above are all consistent with the
qualitative picture of a massive forming star in which the central
object is surrounded by a rotating disk from which it is accreting. A
rotationally-flattened envelope surrounds the central star and its
disk, while the star drives a fast bi-polar wind. 

In this section we present a kinematic analysis of the CO bandhead
emission and the low-J CO absorption lines and use them to derive the
mass of the object about which they rotate.


\begin{figure*}
  \includegraphics[width=17cm]{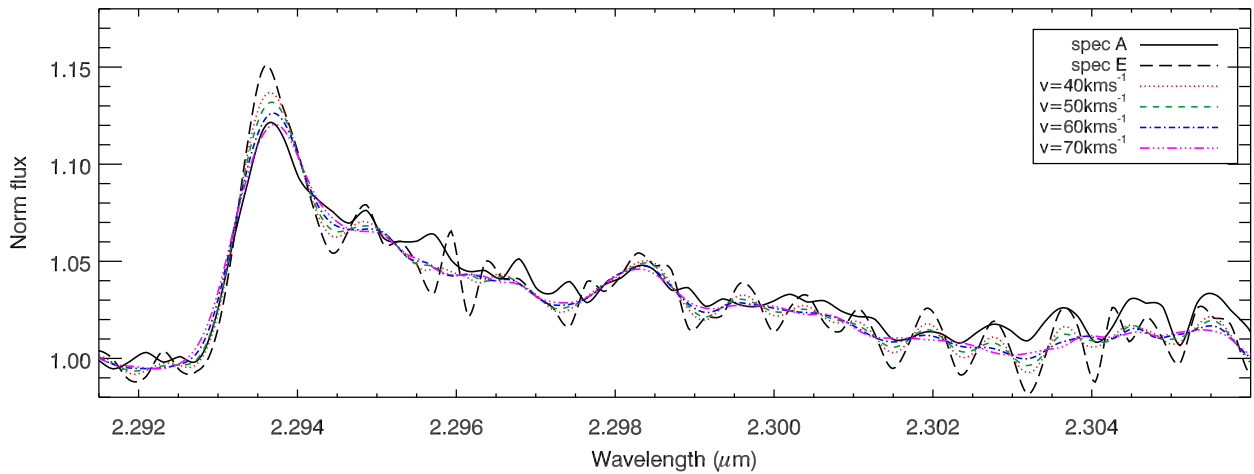}
  \caption{The observed profile of the CO bandhead in region `A'
    (black solid line), as well as the profile seen in the southern
    nebula (from region `C') when convolved with a PSF of various
    widths. }
  \label{fig:coconv}
\end{figure*}

\subsection{The bandhead broadening}

As shown in Fig.\ \ref{fig:coblue}, the CO bandhead emission lines
appear to be broader towards region `A' than those observed in the
southern nebula. This behaviour can be attributed to a flattened
rotating structure (i.e.\ a disk) which is seen at lower inclination
when viewed directly than when seen reflected by the nebula to the
south. If we assume the extreme case that (a) the line-of-sight `A'
views the disk edge-on, and (b) that the southern nebula views the
disk pole-on, by convolving the profile seen from the southern nebula
until it matches that at `A' we can estimate a lower limit to the
disk rotation speed. By assuming Keplerian rotation we can then
estimate the mass of the central object about which the disk rotates.

In Fig.\ \ref{fig:coconv} we plot the CO bandhead profile seen at
`A'. We also plot the profile seen in the southern nebula (region `C')
which has been convolved with a rotational broadening function of the
form $\psi(v) \sim [ 1 - (v/V_{\rm rot})^2]^{1/2}$, where $v$ is
velocity and $V_{\rm rot}$ is the disk rotational velocity. The best
match to the blue-edge of the bandhead, as well as the lower-J line
profiles, is found when the rotational velocity is 50$\pm$5\kms. If
the direct line-of-sight is not exactly edge-on, and/or if the
orientation of the reflected light is not exactly pole-on, the disk
rotation speed will be greater than this value. By taking the
inclination angle recently derived by de~Wit et al.\ (submitted) of
60$\pm$20\degr\ (i.e.\ 30\degr\ away from being edge-on), we can further
constrain the disk rotation speed to 64$^{+22}_{-8}$\kms. 

Assuming that the disk rotation is Keplerian, and that the CO emission
arises in a narrow annulus at a fixed radius $R$ from the
central object we can now derive a mass of the central object $M_{c}$
from,

\begin{equation}
  M_{c} = R V^{2} / G
  \label{equ:kep}
\end{equation}

\noindent where $G$ is the gravitational constant. \citet{Bik04} who
found that the CO typically forms at a distance from the central star
of 1-3AU. Using $R = 2\pm1$AU, and a distance to W33A of 3.8kpc
\citep{Faundez04}, we find that the lower limit to the central mass is
$M_{c} = 10 ^{+9}_{-5}$\msun.


\subsection{The absorption-line velocity structure}
\begin{figure}
  \includegraphics[width=8.5cm]{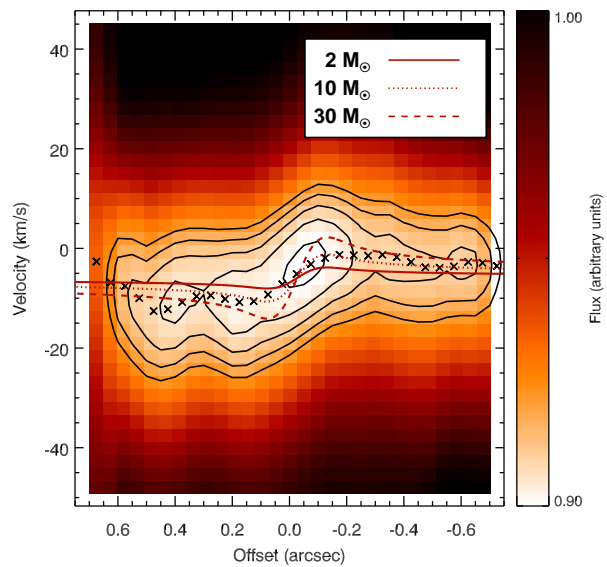}
  \caption{Position-velocity diagram across the rotating structure
    seen in the CO absorption lines. The image is scaled in terms of
    fractional intensity below the continuum. The crosses indicate the
    velocity of the intensity minimum for the absorption as a function
    of positional offset, and contours are drawn at intensity levels
    of 0.895 to 0.92 at intervals of 0.005. Lines show models of
    Keplerian rotation for three different central masses and an
    assumed distance of 3.8kpc. }
  \label{fig:co_pv}
\end{figure}

In Sect.\ \ref{sec:coabs} we showed that the absorption lines of CO
appear to be formed in a rotating structure coincident with region A,
and measured the PA of this structure to be 60$\pm$10\degr. In
\fig{fig:co_pv} we show a position-velocity slice across the averaged
CO datacube at a PA of 60\degr. The data shows the classical signature
of a Keplerian rotating structure. To help illustrate this,
overplotted are models of Keplerian rotation for three different
central masses and with an assumed distance of 3.8kpc
\citep{Faundez04}. Each curve has been convolved with the spatial
resolution of our observations (0.13\arcsec). A formal least-squares
fit to the data yields a central mass of $M \sin i = 13 \pm 2$\msun, a
result which is insensitive to the uncertainty in PA. If we again
assume the system inclination is 60$\pm$20\degr\ we find that the
outer envelope sees a central mass of $15^{+5}_{-3} $\msun. Note that
the central mass seen by the envelope will include the mass of the
envelope itself. This number is very similar to that derived for the
central mass enclosed by the circumstellar disk, and suggests that the
majority of the system's mass is contained within the central star.

If the central object has arrived on the main-sequence, this mass
would correspond to a spectral type earlier than B1 and a zero-age
min-sequence luminosity of $\log (L/$\lsun$ = 4.3 \pm 0.3$
\citep[e.g.][]{Mey-Mae00}. This is in good agreement with the latest
estimation of W33A's bolometric luminosity from MIPS photometry, $\log
(L/$\lsun$ = 4.7$ (Mottram et al.\ 2009, in prep). The derived
spectral-type and luminosity may also explain the detection of the
\mgii\ emission lines. Such lines are typically seen in the spectra of
early-to-mid B supergiants, and are primarily formed in the wind
\citep{Hanson05}. The ionization emission lines that we see in W33A
(\brg\ and \mgii) are probably formed in the bi-polar outflow, the
density of which may be similar to that of a typical B supergiant
wind.

\section{Summary and conclusions} \label{sec:conc}
We have presented a study of adaptive-optics assisted near-infrared
integral-field spectroscopy of the young stellar object W33A.  The
results can be summarized as follows:

\begin{itemize}
\item In the first 3-D spectro-astrometric study of its kind, we have
  resolved the \brg\ emission on sub-milliarcsecond scales (physical
  scales $<$1\,AU), and find that the emission geometry is that of a
  fast bipolar wind which is aligned with the large-scale outflow seen
  in wide-field images. The H$_2$ 2.12\microns\ emission also traces
  an outflow with the same orientation out to $\sim$1\arcsec.
\item The extended nebulosity to the south is seen in the continuum,
  as well as in the light of \brg\ and CO. We have shown that this
  emission is light reflected by the outflow lobe, and so can
  be used as a `mirror' with which to observe the emission from the
  central object from different angles. 
\item We interpret the CO emission as arising in a disk which has a
  low inclination with respect to our line-of-sight. In the southern
  nebula we see light reflected towards us from the disk at a lower
  inclination angle (i.e. more pole-on), which causes the bandhead
  profile to be less velocity-broadened. This is consistent with the
  disk plane being perpendicular to the large-scale outflow. The
  velocity broadnening implies a lower-limit to the central mass of
  $M_{c} = 10 ^{+9}_{-5}$\msun.
\item From the low-J v=2-0 transitions of CO at $\sim$2.35\microns\ we
  find evidence for a rotationally-flattened cool molecular envelope,
  or `torus', at a radius of $\sim$2000\,AU from the central star. The
  plane of rotation is perpendicular to the small- and large-scale
  outflows.  Analysis of the velocity structure of the torus indicates
  that it is rotating about the outflow's axis, and that it is
  orbiting a central mass of $15^{+5}_{-3} $\msun.
\end{itemize}

Our findings suggest a picture of massive star formation within W33A
which is in excellent qualitative agreement with the
accretion-disk-plus-bipolar wind paradigm. An accretion disk orbits a
massive central star, which is surrounded by a cool molecular envelope
which has been rotationally-flattened into a torus. The central star
is driving a bipolar wind, seen on small scales in the ionized gas,
and on larger scales as molecular / continuum emission. The two
measurements of the central mass indicate that at most the accretion
disk makes up $\sim$30\% of the system, and so the system mass is
dominated by the central star.

\section*{Acknowledgments}
Based on observations obtained at the Gemini Observatory, which is
operated by the Association of Universities for Research in Astronomy,
Inc., under a cooperative agreement with the NSF on behalf of the
Gemini partnership: the National Science Foundation (United States),
the Science and Technology Facilities Council (United Kingdom), the
National Research Council (Canada), CONICYT (Chile), the Australian
Research Council (Australia), Ministério da Ciência e Tecnologia
(Brazil) and Ministerio de Ciencia, Tecnología e Innovación Productiva
(Argentina).

\bibliographystyle{/fat/Data/bibtex/apj}
\bibliography{/fat/Data/bibtex/biblio}

\appendix

\section{Spectro-astrometry of the telluric standard star}
\begin{figure}
  \centering
  \includegraphics[width=8.5cm,bb=0 0 580 510]{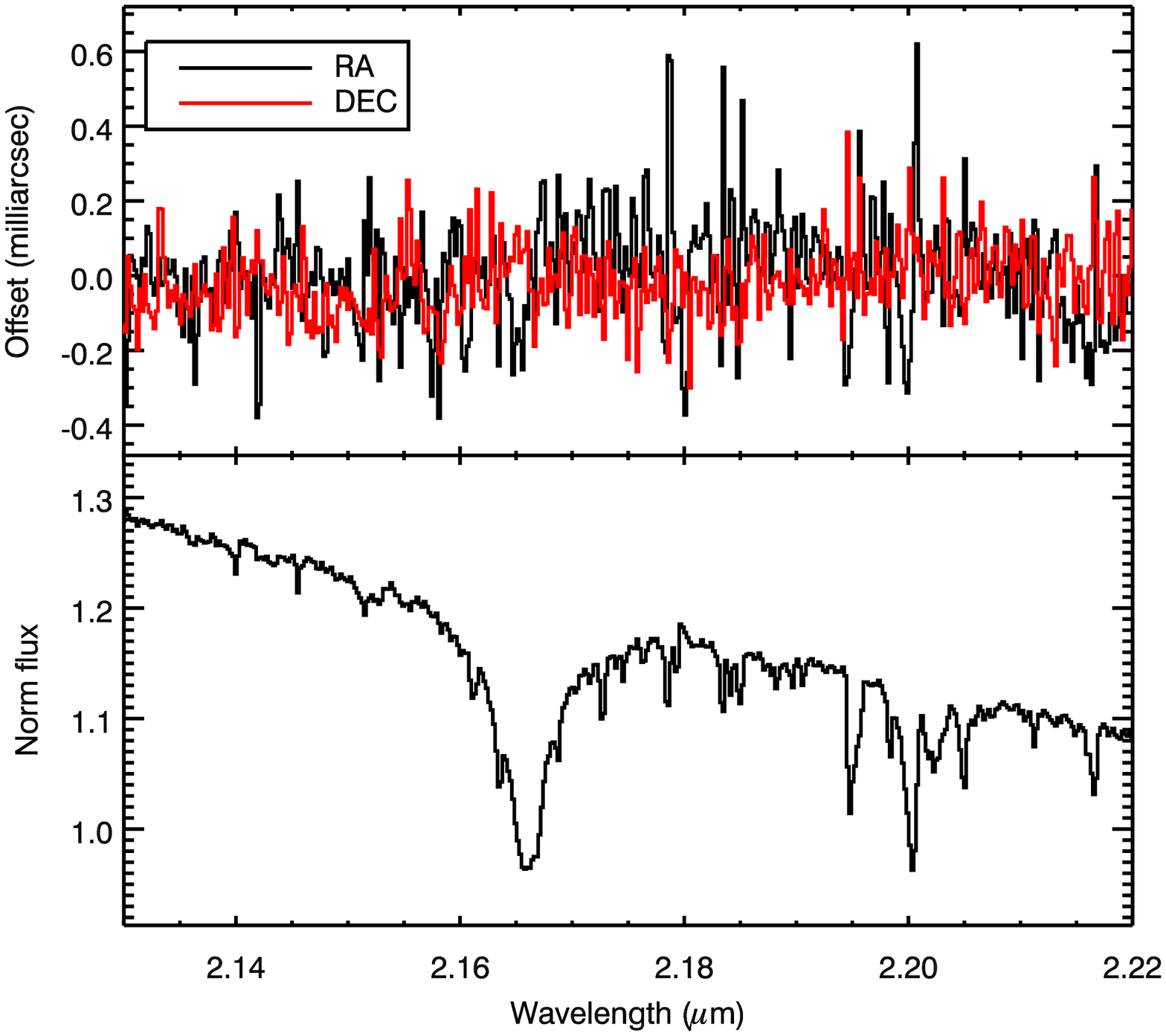}
  \caption{Spectro-astrometry of the \brg\ line for the telluric
    standard star used in our observations, similar to
    Fig.\ \ref{fig:specast}. The lack of artifacts across the telluric
    features and the star's \brg\ absorption, as well as the observed
    level of astrometric precision ($\sim$0.2mas), validate the
    results of the similar analysis of W33A. }
  \label{fig:specast_tell}
\end{figure}

In Fig.\ \ref{fig:specast_tell} we show a spectro-astrometric analysis
of the telluric standard star following a similar methodology to that
presented in Sect.\ \ref{sec:specast}. The intensity spectrum (bottom
panel) shows the \brg\ absorption line, as well as the many telluric
features. The top panel shows the astrometry of the star's flux peak
as a function of wavelength, illustrating that no discernable
variations are seen across neither the \brg\ line nor the telluric
lines. This of course is to be expected, and the Figure serves to show
that the technique we employ produces no spurious artifacts greater
than the noise level ($\sim$0.2mas) which could be misinterpreted. 

\end{document}